\documentstyle[aps,prd,epsf,floats,axodraw,graphicx]{revtex}
\tighten
\begin{document}

\preprint{\makebox{\begin{tabular}{r} 
 OUTP-99-59-P\\ hep-ph/9911365 \\ \end{tabular}}}
\title{A supersymmetric solution to the KARMEN time anomaly}
\author{\large Debajyoti Choudhury$^1$, Herbi Dreiner$^2$, 
            Peter Richardson$^3$, Subir Sarkar$^3$ \\}
\address{
$^1$ Mehta Research Institute, Chhatnag Road, Jhusi, Allahabad 211019, INDIA\\
$^2$ Rutherford Appleton Laboratory, Chilton, Didcot OX11 0QX, UK\\
$^3$ Theoretical Physics, University of Oxford, 1 Keble Road,
      Oxford OX1 3NP, UK}
\date{\today}
\maketitle
\begin{abstract}
We interpret the KARMEN time anomaly as being due to the production of
a (dominantly bino) neutralino with mass 33.9 MeV, which is the
lightest supersymmetric particle but decays into 3 leptons through the
violation of R-parity. For independent gaugino masses $M_1$ and $M_2$
we find regions in the ($M_1$, $M_2$, $\mu$, $\tan\beta$) parameter
space where such a light neutralino is consistent with all
experiments. Future tests of this hypothesis are outlined.
\end{abstract}

\section{Introduction}

In 1995, the KARMEN experiment at the Rutherford Appleton Laboratory
reported an anomaly in the time distribution of the charged and
neutral current events induced by neutrinos from $\pi^+$ and $\mu^+$
decays at rest \cite{karmen}. This was ascribed to the production of a
new particle, denoted $x$, in the anomalous pion decay
\begin{equation} 
 \pi^+ \rightarrow \mu^+ + x, 
\label{piondec} 
\end{equation}
with a small branching ratio in the range $\sim10^{-16}-10^{-8}$
depending on the lifetime of $x$. The particle must be neutral since
it passes through over 7~m of steel shielding. The time-of-flight to
the detector is $3.6\pm0.25\,\mu{\rm s}$, implying that the particle
moves non-relativistically with velocity $v_x=5.2^{+2.2}_{-1.4}
\times10^6$~m\,s$^{-1}$. This requires its mass to be 33.9~MeV and its
kinetic energy to be $T_x\approx5$~keV.\footnote{The required mass is
within $0.02\%$ of the pion--muon mass difference.} The observed
energy in the detector is $\sim11-35$~MeV, which must therefore come
from the decay of $x$. Since 1995 the KARMEN experiment has been
upgraded to significantly reduce the cosmic ray background. It has
recently been reported that the time anomaly persists in the new data
\cite{win}.  A time-of-flight likelihood analysis adopting the
hypothesis that it is due to a decaying particle as described above
has a negative natural log-likelihood ratio of 9, i.e. less than 1 in
$10^4$ chance of being a statistical fluctuation. The significance is
thus sufficiently high that we are motivated to reexamine its physical
origin.

There have already been several proposals to explain the KARMEN
anomaly \cite{barger,choudhury,krasnikov}. In Ref.\cite{barger}, the
authors considered in detail the possibility that $x$ is a neutrino
and concluded that a $SU(2)_{\rm L}$ doublet neutrino was excluded by
existing data. This was further reinforced by the subsequent
improvement \cite{daum} in the experimental upper limit on the
branching ratio,
\begin{equation}
 {\rm BR} (\pi^+ \rightarrow \mu^+ + x) < 1.2 \times 10^{-8} 
 \quad {\rm (95\%~C.L.)}, 
\label{psi}
\end{equation}
versus the minimum value of $\sim2\times10^{-8}$ required in the
doublet neutrino interpretation. However a sterile neutrino
interpretation was found to be consistent, within strict limits on the
mixing parameters (see also Ref.\cite{govaerts}), although this may
still be in conflict with astrophysical and cosmological constraints
\cite{barger}.

In Ref.\cite{krasnikov} a solution was proposed based on the anomalous
muon decay $\mu^+\rightarrow\,e^++x$, where $x$ is taken to be a
scalar boson of mass 103.9~MeV. However this implies too large a value
for the energy released in the $x$ decay and the required branching
ratio is also constrained by the recent bound
BR$(\mu^+\rightarrow\,e^++x)<5.7\times10^{-4}$ (90\%~C.L.)
\cite{bilger}. Thus it is necessary to add to the model 2 other
scalar bosons into which $x$ can cascade decay in order to dilute the
energy \cite{krasnikov}. This model appears viable but is somewhat
baroque.

In Ref.\cite{choudhury}, a supersymmetric solution was considered. The
$x$ particle was interpreted as a photino (or zino) and the anomalous
pion decay
\begin{equation}
 \pi^+ \rightarrow \mu^+ + \tilde\gamma
\end{equation} 
was assumed to proceed via the R-parity violating operator
$L_2Q_1D^c_1$.~\footnote{For reviews on R-parity violation see
Ref.\cite{dreiner}.} The same operator then enables the photino to
decay radiatively as
\begin{equation}
 \tilde\gamma \rightarrow \gamma + \nu_\mu
\end{equation}
via a one-loop diagram with a $d$ quark and $\tilde{d}$ squark in the
loop. However the expected peak at 17~MeV has not been reported in the
new data \cite{win} on the energy spectrum of the anomalous events, so
a 2-body decay for the $x$ particle seems disfavoured. Therefore this
model \cite{choudhury} may not be viable in its present form. We
present below the necessary extension to produce a 3-body decay for
such a light neutralino.
\footnote{Such a decay was also proposed in Ref.\cite{apostolos} which
invoked possible mixing between neutrinos and gauginos/higgsinos as
the reason for neutralino instability rather than R-parity violating
vertices. However to explain the KARMEN anomaly then requires the
Higgs mixing term $\mu\,H_1\,H_2$ in the superpotential to be
unnaturally small, $\mu\leq30$~MeV. Moreover this scenario implies a
MeV mass $\nu_{\tau}$ which is definitively ruled out by cosmological
and astrophysical arguments \cite{bbn,raffelt}.}

\section{The Model}

We consider the lightest neutralino in supersymmetry,
$\tilde\chi^0_1$, to be the hypothetical $x$ particle, with mass
$m_{\tilde\chi^0_1}=33.9$~MeV. This will also be the lightest
supersymmetric particle (LSP) in our model. Since $x$ is effectively
stable on collider time-scales (this is quantified below) our model
will experimentally look very similar to the MSSM. Now in a
GUT-inspired MSSM, $M_1=(5/3)\tan^2\theta_{\rm W}M_2$, and assuming
this relation requires $m_{\tilde\chi^0_1}>32.3$~GeV from current LEP
data \cite{aleph}. Thus in order to obtain a very light neutralino we
must consider $M_1$ and $M_2$ to be independent parameters. A small
$M_2$ implies at least one light chargino which is excluded by
experiment, while a small $M_1$ implies that the LSP will be
dominantly bino. We will quantify this below and determine regions in
the $(M_1,M_2,\mu,\tan\beta)$ parameter space consistent with all
experimental limits. The solutions indeed turn out to be dominantly
bino with a small higgsino contribution.

\begin{figure}[h]
\centering
\bigskip
\includegraphics[width=10cm]{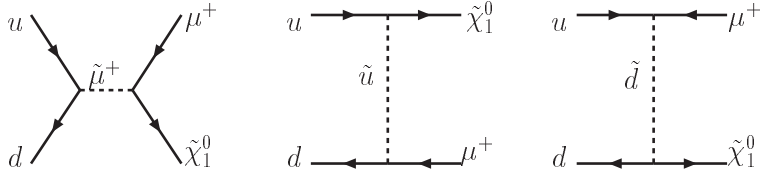}
\bigskip
\caption{Tree-level Feynman diagrams for pion decay via the
 operator $L_2Q_1D^c_1$.}
\label{fig:pion}
\end{figure}

Furthermore we invoke 2 non-zero R-parity violating operators. The
pion decay
\begin{equation}
 \pi^+ \rightarrow \mu^+ \tilde\chi^0_1
\label{piondec2}
\end{equation}
proceeds through the operator $\lambda'_{211}L_2Q_1D^c_1$ and the
leading order Feynman diagrams are shown in Fig.~\ref{fig:pion}. The
neutralino is assumed to decay as
\begin{equation}
 \tilde\chi^0_1 \rightarrow e^+ e^- \nu_{\mu,\tau}
\label{neutdec}
\end{equation}
through either $\lambda_{121}L_eL_{\mu}E^c_e$ or $\lambda_{131}L_e
L_{\tau}E^c_e$. Note that this is the only kinematically accessible
tree-level 3-body visible decay for such a light LSP.
%
%

\begin{figure}[t]
\centering
\bigskip
\includegraphics[angle=90,width=10cm]{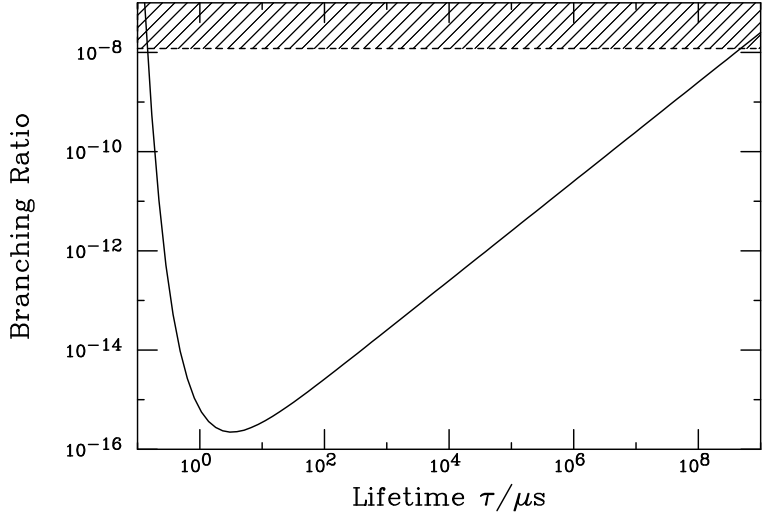}
\bigskip
\caption[dummy]{The solutions to the KARMEN anomaly in terms of the
 anomalous pion branching ratio and the $x$-particle lifetime. The
 hashed area denotes the experimental upper bound
 (\protect\ref{psi}).}
\label{fig:karmensol}
\end{figure}

In Fig.~\ref{fig:karmensol} we show the values of the branching ratio
for $\pi^+\rightarrow\mu^++\tilde\chi^0_1$ and lifetimes
$\tau_{\tilde\chi^0_1}$ which are compatible with the KARMEN data
\cite{karmen,win}.  In order to determine the required range of the
couplings $\lambda'_{211}$, $\lambda_{1\{2,3\}1}$ in our model which
are consistent with the solutions in Fig.~\ref{fig:karmensol}, we must
first determine the pion branching ratio in terms of the
supersymmetric parameters. The partial width as computed from the
diagrams in Fig.~\ref{fig:pion} is
\begin{eqnarray}
 \Gamma(\pi \rightarrow \mu {\tilde\chi}^0_1) &=&
 \frac{{\lambda'}^2_{211} f^2_\pi m^2_\pi p_{\rm cm}}{8\pi(m_d + m_u)^2}
 \left(\frac{A_e}{M^2_{\tilde\mu}} - \frac{A_u}{2M^2_{\tilde{u}}} -
 \frac{A_d}{2M^2_{\tilde{d}}}\right)^2
 \left(m^2_\pi - m^2_\mu - m^2_{\tilde\chi^0_1}\right),
\end{eqnarray}
where $m_\pi$ and $m_\mu$ denote the pion and the muon masses,
$M_{\tilde\mu,\tilde{u},\tilde{d}},$ denote the corresponding scalar
fermion masses, $m_u,\, m_d$ are the first generation {\it current}
quark masses, and $f_\pi$ is the charged pion decay constant. The
constants $A_{e,u,d}$ refer to the neutralino coupling and are given
in Table~\ref{tab:picp} both for the general case and for the limiting
cases of either a pure bino or photino neutralino. The phase space
factor is given by $p_{\rm cm}=([m^2_\pi-(m_\mu+m_{\tilde\chi^0_1})^2]
[m^2_\pi-(m_\mu-m_{\tilde\chi^0_1})^2])^{1/2}/(2m_\pi)$. In the
Appendix we give some details of how this result is obtained.

%
%

%
%
\begin{table}[b]
\caption{Neutralino coupling coefficients for the pion decay.}
\begin{tabular}{cccc}
 Coefficient & General formula & Pure photino & Pure bino \\
\tableline
 $A_e$  & $e N'_{l1} + \frac{gN'_{l2}}{\cos\theta_{\rm
  W}}\left(\frac{1}{2}-\sin^2\theta_{\rm W}\right)$
        & $e$
        & $-g' Y_{e{\rm L}}$\\
\hline
 $A_u$  & $ -e e_u N'_{l1} - \frac{gN'_{l2}}{\cos\theta_{\rm W}}\left(
  \frac{1}{2}-e_u\sin^2\theta_{\rm W}\right)$
        & $ -e e_u $
        & $-g' Y_{u{\rm L}}$ \\
\hline
 $A_d$  & $ e e_d N'_{l1} - \frac{g e_d \sin^2\theta_{\rm W} 
  N'_{l2}}{\cos\theta_{\rm W}}$
        & $e e_d$
        & $g' Y_{d_{\rm R}}$ \\
\end{tabular}
\label{tab:picp}
\end{table}

The branching ratio for the anomalous pion decay, assuming the decay
$\pi^+\rightarrow\mu^+\nu_\mu$ to be dominant, is given
by~\footnote{This corrects the result given earlier \cite{choudhury}.}
\begin{eqnarray} 
 {\rm BR}
 (\pi \rightarrow \mu {\tilde\chi}^0_1) &=& \frac{{\lambda'}^2_{211} m_\pi^5
 p_{\rm cm}}{2G_{\rm F}^2 m_\mu^2 (m_d + m_u)^2}
 \left(\frac{A_e}{M^2_{\tilde\mu}}-\frac{A_u}{2M^2_{\tilde{u}}}-
 \frac{A_d}{2M^2_{\tilde{d}}}\right)^2 \frac{\left(m^2_\pi - m^2_\mu
 - m^2_{\tilde\chi^0_1}\right)} {(m_\pi^2 - m_\mu^2)^2} \\ \nonumber
 &\approx &2.6 \times 10^{-8}\;
 \left(\frac{{\lambda'}_{211}}{10^{-4}}\right)^2
 \left(\frac{150\mbox{~GeV}}{M_{\tilde f}}\right)^4
\label{pionbr} \\ 
 &< &1.2 \times 10^{-8}.
\label{psi2}
\end{eqnarray} 
To obtain a numerical estimate, we have assumed in Eq.~(\ref{pionbr})
that the scalar fermions are mass degenerate, $M_{\tilde\mu,
\tilde{u},\tilde{d}}=M_{\tilde f}$, and that the neutralino is pure
bino. In the last line we have quoted the experimental bound
(\ref{psi}), shown as a hashed area in Fig.~\ref{fig:karmensol}. This
bound can be satisfied by a small coupling and/or a large sfermion
mass. It can also be satisfied by a fine-tuned cancellation between
different diagrams for distinct sfermion masses, but we disregard this
possibility. The last inequality (\ref{psi2}) can be translated into
an upper bound on $\lambda'_{211}$:
\begin{equation}
 \lambda'_{211} < 6.8 \times 10^{-5}
 \left(\frac{M_{\tilde f}}{150\mbox{~GeV}}\right)^2.
\label{psi3}
\end{equation}

%
%
\begin{table}
\caption{Coefficients for the neutralino decay}
\begin{tabular}{cccc}
 Coefficient & General formula & Pure photino & Pure bino \\
\tableline
 $B_1$ & 
   $-\left( eN'_{l1} + \frac{gN'_{l2}}{\cos\theta_{\rm W}} 
   \left[\frac{1}{2} - \sin^2\theta_{\rm W}\right]\right)$
   & $-e$  & $g'Y_{e{\rm L}}$ \\
\hline
 $B_2$ & $\frac{gN'_{l2}}{2 \cos\theta_{\rm W}}$ & 0 &$g'Y_{e{\rm L}}$ \\
\hline
 $B_3$ &
   $\left(eN'_{l1} - 
   \frac{gN'_{l2}\sin^2\theta_{\rm W}}{\cos\theta_{\rm W}}\right)$ & $e$ 
   & $-g'Y_{e_{\rm R}}$ \\
\end{tabular}
\label{tab:lspcpl}
\end{table}

In the limit where $M_{\tilde f}\gg m_{{\tilde\chi}^0_1}$, the
neutralino decay rate for the operator $\lambda_{1j1}L_1L_jE_1^c$,
$j=2,3$, is given by \cite{dawson}
\begin{eqnarray} 
\Gamma ({\tilde\chi}^0_1 \rightarrow
 e^+ \bar\nu_j e^-) &=& \frac{\lambda^2_{1j1} m^5_{{\tilde\chi}^0_1}}
  {3072 \pi^3} \left(\frac{B^2_1}{M^4_{{\tilde e}{\rm L}}}
  + \frac{B^2_2}{M^4_{{\tilde\nu}^j{\rm L}}} 
  + \frac{B^2_3}{M^4_{{\tilde e}_{\rm R}}} 
  - \frac{B_1B_2}{M^2_{{\tilde e}{\rm L}}M^2_{{\tilde\nu}^j{\rm L}}}
  - \frac{B_1B_3}{M^2_{{\tilde e}{\rm L}}M^2_{{\tilde e}_{\rm R}}}
  - \frac{B_2B_3}{M^2_{{\tilde\nu}^j{\rm L}}M^2_{{\tilde e}_{\rm R}}}
  \right)\nonumber\\ 
  &=& \frac{3\alpha\lambda^2_{1j1} m^5_{\tilde\chi^0_1}}{1024\pi^2
      \cos^2\theta_{\rm W} M_{\tilde f}^4},
\end{eqnarray} 
where $M_{{\tilde e}{\rm L},{\tilde e}_{\rm R},{\tilde\nu}^j{\rm L}}$
denote the scalar lepton masses, and $B_{1,2,3}$ are the relevant
${\tilde \chi}^0_1f{\tilde f}$ couplings given in
Table~\ref{tab:lspcpl}. In the second equation we have again assumed a
pure bino LSP and degenerate scalar fermions. As a numerical estimate
for the lifetime of the bino LSP with $m_{{\tilde\chi}^0_1}=33.9$~MeV
we obtain
\begin{eqnarray} 
 \tau_{\rm bino} &=& 13.2\,{\rm s}
  \left(\frac{0.01}{\lambda_{1\{2,3\}1}}\right)^2
  \left(\frac{M_{\tilde{f}}}{150\mbox{~GeV}}\right)^4
 \label{lifetime}\\ 
 &< &4.78\times10^2\mbox{~s},
\label{psi4} 
\end{eqnarray} 
where the last inequality is obtained by
using the bound (\ref{psi}) and the set of solutions shown in
Fig.~\ref{fig:karmensol}. The resulting bound on the coupling is
\begin{equation}
\lambda_{1\{2,3\}1} > 1.66 \times 10^{-3} 
 \left(\frac{M_{\tilde{f}}}{150\mbox{~GeV}}\right)^2 .
\label{psi5}
\end{equation}
Given the perturbative upper bound, $\lambda_{ijk}<\sqrt{4\pi}$, and
the lower bound on the sfermion mass from LEP 2,
$M_{\tilde{f}}>100$~GeV, we also have a lower limit on the lifetime of
$\tau_{\rm bino}>2.6\times10^{-4}$~s. Thus there is a solution range
of 6 orders of magnitude in lifetime or 3 orders of magnitude in
coupling. For these lifetimes the LSP is stable on collider physics
time scales.

%
%
\begin{figure}
\centering
\bigskip
\includegraphics[angle=90,width=10cm]{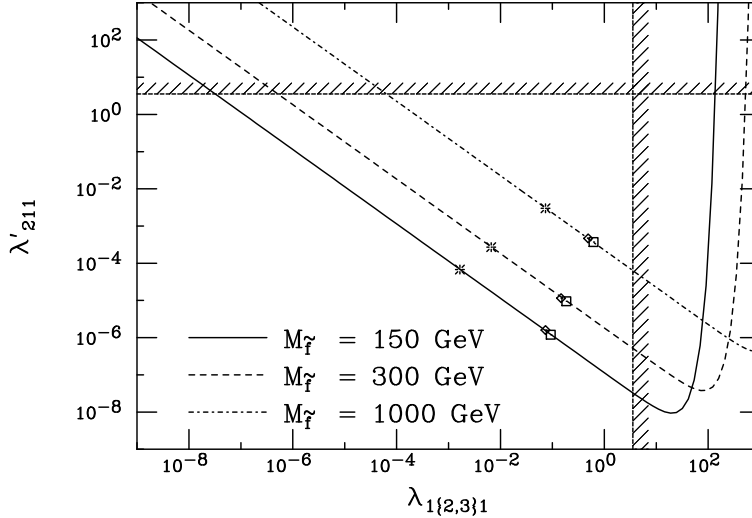}
\caption[dummy]{Solutions to the KARMEN anomaly in terms of the
 R-parity violating couplings $\lambda'_{211}L_2Q_1D^c_1$ and
 $\lambda_{1\{2,3\}1}$, for different (assumed degenerate) sfermion
 masses. The hashed lines indicate upper limits on the couplings from
 perturbativity. The stars and diamonds (squares) give the upper
 limits on the couplings $\lambda'_{211}$ and $\lambda_{121}$
 ($\lambda_ {131}$), respectively. Solutions above and to the left of
 the stars are excluded, as are solutions below and to the right of
 the squares (diamonds).}
\bigskip
\label{fig:oursols}
\end{figure}

We now have all the ingredients to fix the model parameters. In our
model, each point along the curves in Fig.~\ref{fig:karmensol}
corresponds to a specific anomalous pion branching ratio
(\ref{pionbr}) and a specific neutralino lifetime (\ref{lifetime}). If
we assume the scalar fermions are mass degenerate, we can translate
this into specific values of $\lambda'_{211}$ and
$\lambda_{1\{2,3\}1}$ for a fixed sfermion mass. This set of solutions
in the R-parity violating parameter space is shown in
Fig.~\ref{fig:oursols} for $M_{\tilde f}=150$~GeV (solid line),
$M_{\tilde f}=300$~GeV (dashed line), and $M_{\tilde f}=1000$~GeV
(dot-dashed line). The hashed lines at $\lambda, \lambda' =
\sqrt{4\pi}$ denote the perturbative limit. For large scalar fermion
masses ($>1$~TeV) we quickly run out of room for perturbative
solutions. The solutions above and to the left of the stars are
excluded by the inequalities (\ref{psi3}, \ref{psi5}).

\section{Constraints on the R-parity Violating Couplings}\label{sec:limits}

The R-parity violating couplings we have introduced violate lepton
number and are thus constrained by laboratory experiments. The best
bounds at the $2\sigma$ level have been been summarized as
\cite{dedes}
\begin{eqnarray}
 \lambda'_{211} &< & 0.059\; 
 \left( \frac{M_{{\tilde d}_{\rm R}}}{100\mbox{~GeV}}\right) , \nonumber\\
 \lambda_{121}  &< & 0.049 \;
 \left(\frac{M_{{\tilde e}_{\rm R}}}{100\mbox{~GeV}}\right)
 \quad \Rightarrow \tau_{\rm bino} > 0.24 \mbox{~s},
\label{bounds}\\
 \lambda_{131}  &< & 0.062 \;
 \left(\frac{M_{{\tilde e}_{\rm R}}}{100\mbox{~GeV}}\right)
 \quad \Rightarrow \tau_{\rm bino} > 0.15 \mbox{~s}. \nonumber
\end{eqnarray}
The bound on $\lambda'_{211}$ is from measurements of $R_\pi=\Gamma
(\pi\rightarrow e\nu)/ \Gamma(\pi\rightarrow\mu\nu)$ \cite{bgh}, the
bound on $\lambda _{121}$ is from charged-current universality
\cite{bgh}, while the bound on $\lambda_{131}$ is from a measurement
of $R_\tau=\Gamma(\tau\rightarrow e\nu{\bar\nu})/\Gamma(\tau
\rightarrow\mu \nu{\bar\nu})$ \cite{bgh}. The above bound on $\lambda
'_{211}$ is weaker than the bound (\ref{psi3}), and we do not consider
it further. In Fig.~\ref{fig:oursols} the above bounds on the coupling
$\lambda_{121}$ and $\lambda_{ 131}$ forbid solutions to the right of
the diamonds and squares, respectively. We are thus left with a range
of solutions of about 2 orders of magnitude in $\lambda'_{2 11}$ and
$\lambda_{1\{2,3\}1}$. This corresponds to 4 orders of magnitude in
the pion branching ratio and in the LSP lifetime, respectively. In the
last 2 equations we have translated the upper bound on
$\lambda_{1\{2,3\}1}$ into a lower bound on the lifetime using
Eq.~(\ref{lifetime}), to be compared with the upper bound
(\ref{psi4}). Note that these bounds are independent of the sfermion
mass.

Besides bounds on individual couplings, we must also consider bounds
on the product of the couplings $\lambda'_{211}\lambda_{121}$ or
$\lambda'_{211} \lambda_{131}$. In the first case, we can get an
additional contribution to pion decay
$\pi^+\rightarrow{\tilde\mu}^+\rightarrow e^+\nu_e$ which changes the
prediction for $R_\pi$
\begin{equation}
 R_\pi = R_\pi^{\rm SM} \left[1-\frac{m^2_\pi{\lambda'}_{211}\lambda_{121}}
         {2\sqrt{2}G_{\rm F} M^2_{\tilde{\mu}{\rm L}} m_e (m_u+m_d)}
	 \right]^2.
\end{equation}
As the corresponding Feynman diagram has a different structure from
the t-channel squark exchange which gives the bound on
${\lambda'}_{211}$ in Eq.~(\ref{bounds}) we get a much stricter bound
on the product of the couplings than on either of the couplings
individually. This leads to the following bound at the $2\sigma$ level
\begin{equation}
 \lambda'_{211}\lambda_{121} < 4.6 \times 10^{-7} \;
 \left(\frac{m_{{\tilde \mu}{\rm L}}}{100\mbox{~GeV}}\right)^2.
\end{equation}
This means that in the case of $\lambda_{121}$ (as opposed to
$\lambda_{131}$) the maximum scalar fermion mass which will solve the
KARMEN anomaly in our model is 450~GeV.

The couplings $\lambda'_{211}$ and $\lambda_{131}$ violate muon and 
tau lepton number, respectively, and can thus lead to the decay
$\tau\rightarrow\mu\gamma$. The experimental bound has recently been
improved \cite{cleo},
\begin{equation}
 {\rm BR}(\tau \rightarrow \mu \gamma) < 1.0 \times 10^{-6} \quad 
                                         {\rm (90\%~C.L.)},
\end{equation}
but is still 4 orders of magnitude weaker than the experimental
upper bound on BR($\mu\rightarrow\,e\gamma$). Therefore a bound on a
product of couplings which yield the decay $\tau\rightarrow \mu\gamma$
via a one-loop penguin diagram, e.g. $\lambda_{121} \lambda_{131}$,
must be 2 orders of magnitude weaker than the corresponding bound on
the couplings which give $\mu\rightarrow\,e \gamma$, i.e. one would
expect $\lambda_{121}\lambda_{131}<{\cal O} (10^{-2})$
\cite{huitu}. In our model, the couplings $\lambda'_{211}
\lambda_{131}$ only contribute to the decay $\tau\rightarrow\mu
\gamma$ at the 2-loop level and the bound is thus significantly weaker
than ${\cal O}(10^{-2})$. (The decay $\tau\rightarrow\mu ee$ is
similarly suppressed.) This is significantly weaker than the bound
(\ref{bounds}) so we have no new bounds on the product
$\lambda'_{211}\lambda_{131}$. Furthermore since the bound
(\ref{psi3}) on $\lambda'_{211}$ is so restrictive in our model, we
need not worry about the model dependent bounds from flavour changing
neutral currents \cite{agashe}.

In Refs.\cite{ellis1,sphal,ellis2} severe cosmological bounds were
derived on all R-parity violating couplings from considerations of
GUT-scale lepto/baryogenesis in the early universe:
\begin{equation}
 \lambda,\;\lambda',\;\lambda''< 5 \times 10^{-7} 
 \left(\frac{m_{\tilde f}}{1\,\mbox{~TeV}}\right). 
\label{cosmobound}
\end{equation}
Subsequently it was shown that it is sufficient for just one
lepton-flavour to satisfy this bound \cite{sphal,ellis2}. In
Fig.~\ref{fig:oursols} we can see that for our model both couplings
violate the bound (\ref{cosmobound}). For the case
$(\lambda'_{211},\lambda_{121})$ we must therefore demand that {\em
either} all electron number violating couplings or all tau number
violating couplings satisfy Eq.~(\ref{cosmobound}), while for the case
$(\lambda'_{211},\lambda_{131})$, we must demand that {\em all}
electron number violating couplings satisfy
Eq.~(\ref{cosmobound}). Alternatively, baryogenesis could plausibly
occur at the electroweak scale, in which case the bounds
(\ref{cosmobound}) do not apply.

\section{Experimental Constraints on a Light Neutralino}

We now summarize relevant experimental constraints on a light
neutralino LSP and show that these are satisfied in regions of
$(M_1,M_2,\mu,\tan{\beta})$ parameter space for a dominantly bino
${\tilde\chi}^0_1$ with a small higgsino contribution. In our model,
$M_1$ and $M_2$ are {\em not} related by the supersymmetric grand
unified relation and we treat them as separate free parameters.

\subsection{Bounds from $e^+e^-\rightarrow\nu{\overline\nu}\gamma$}

The process $e^+e^-\rightarrow\nu{\overline\nu}\gamma$ can be measured
in electron-positron collisions by detecting the photon and the
missing energy due to the neutrinos \cite{nunugam}.  As the lightest
neutralino in our model is long-lived on the time scale of
collider experiments, the process $e^+e^-\rightarrow \tilde{\chi}^0_1
\tilde{\chi}^0_1\gamma$ will give the same experimental signature.

\begin{figure}
\centering
\includegraphics[angle=90,width=8cm]{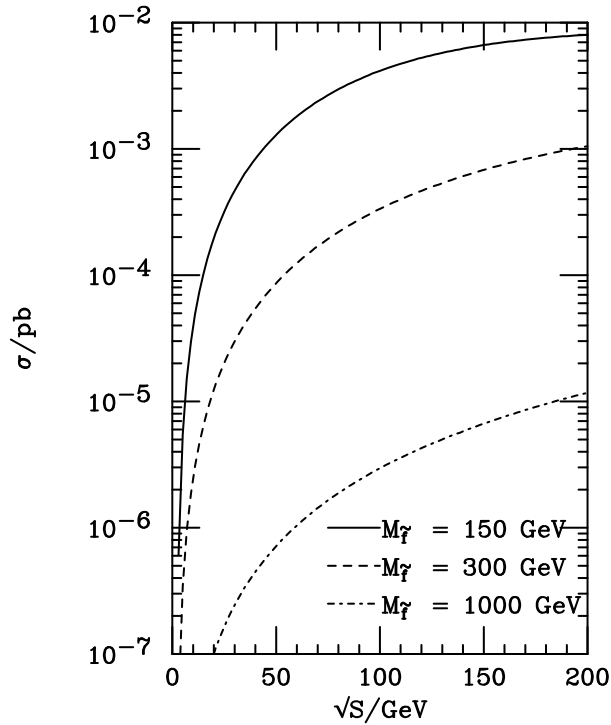}
\caption[dummy]{Cross-section for for the production of a purely bino
 neutralino with mass 33.9~MeV through
 $e^+e^-\rightarrow\tilde{\chi}^0_1\tilde{\chi}^ 0_1\gamma$.}
\label{fig:photon}
\end{figure}

The cross section for this latter process has been calculated
\cite{photino} for the case of a pure photino and can be easily
extended to the pure bino case we are considering here. The cross
section is shown as a function of the centre-of-mass energy in
Fig.~\ref{fig:photon} and is rather low. The expected number of events
for a number of different experiments are given in
Table~\ref{tab:photon} assuming a scalar fermion mass of
$M_{\tilde{f}}=150$GeV. We have used the same cuts on the energy and
angle of the photon as in Ref.\cite{photino}.

%
%

As can be seen from Table~\ref{tab:photon}, {\em no} limits on this
process can be set by LEP, as the expected number of events is much
less than one. The recent results from OPAL \cite{opal} give 138
observed events (with a statistical error of $\pm11.9$) against the
standard model expectation of $141.1\pm1.1$ events from
$e^+e^-\rightarrow\nu{\bar\nu}\gamma$. There is also an expected non-physics
background of $2.3\pm1.1$ events. Thus there is no evidence for any
excess.

%
%
\begin{table}[b]
\caption{Cross-sections for the production of
 $\tilde{\chi}^0_1\tilde{\chi}^0_1\gamma$ at $e^+e^-$ colliders for
 (an assumed common) sfermion mass $M_{\tilde f}=150$~GeV and general
 expectations of the integrated luminosity.}
\begin{tabular}{ccccc}
 Experiment & Integrated luminosity (pb$^{-1}$) & Energy 
  & Cross-section (fb) & Number of events \\
\tableline
LEP    & 	6.65	& 130  & 	$5.87$		& 0.04\\
       & 	5.96	& 136  & 	$6.14$		& 0.04\\
       &	9.89	& 161  &  	$7.11$		& 0.07\\
       &	10.28	& 172  &  	$7.44$		& 0.08\\
       &	54.5	& 183  &  	$7.72$		& 0.42 \\
       &	75.	& 200  & 	$8.05$		& 0.60 \\
\hline
 KEK-B & $1\times10^{5}$&  10.5 & $6.74\times10^{-2}$ 	& 6.7 \\
\hline
 BaBar & $3\times10^4$	& 10.5  & $6.74\times10^{-2}$ 	& 2.0 \\
\hline
 NLC   & $3\times10^5$  & 500   & 6.19                  & 1857
\end{tabular}
\label{tab:photon}
\end{table}

With the higher luminosities expected at the B-factories KEK-B and
BaBar, a few events may be expected. The Standard Model (SM) cross
section at this energy is 2.3~fb, corresponding to $230\pm15$ events
at KEK-B and $70\pm8$ events at BaBar.~\footnote{This corresponds to
one year of running based on the luminosities given in
Ref.\cite{pdg}.}  The statistical uncertainty still exceeds the signal
rate so we do not expect any sensitivity to a light neutralino.

At the NLC we expect a substantially higher number of events. The SM
cross section for the same cuts is about 0.35~pb for 3 neutrinos
\cite{ambrosanio} corresponding to $1.1\times10^5$ events, with a
small statistical error of $330$ events. Thus this can provide a test
of our model.

\subsection{Bounds from the invisible decay of the $Z^0$}

In our model, $m_{{\tilde\chi}^0_1}\ll M_{Z^0}/2$, therefore the decay
$Z^0\rightarrow{\tilde\chi}^0_1{\tilde \chi}^0_1$ is kinematically
accessible, and the ${\tilde\chi}^0_1$ can be considered to be
effectively massless, just like a neutrino. The LSP decays outside the
LEP detectors, thus the process
$Z^0\rightarrow{\tilde\chi}^0_1{\tilde\chi}^0_1$ will contribute to
the invisible width of the $Z^0$. The current measurement of the
invisible $Z^0$ width translated into the number of light neutrino
species is \cite{pdg}
\begin{equation} 
 N_\nu = 3.00 \pm 0.08,  
\end{equation} 
so we must require that $\Gamma(Z^0\rightarrow{\tilde\chi}^0_1{\tilde
\chi}^0_1)<0.08\;\Gamma(Z^0\rightarrow\nu{\bar\nu})$, where the rhs
refers to one neutrino species only.

A pure bino LSP does not couple to the $Z^0$ at tree-level. The
dominant contribution will thus come from the Higgsino admixtures of
the LSP, $N_{13},\;N_{14}$, in the notation of
Ref.\cite{haberkane}. This enters with the fourth power in the decay
rate $Z^0\rightarrow{\tilde\chi}^0_1{\tilde\chi}^0_1$. The Higgsino
has equal strength coupling to the $Z^0$ compared to a neutrino, thus
yielding the constraint
\begin{equation}
\sqrt{|N_{13}|^2+|N_{14}|^2}<0.5 \approx (0.08)^{1/4} .
\label{higgs}
\end{equation}
We shall see below that it is straightforward to find regions which
satisfy this in $(M_1,M_2,\mu,\tan\beta)$ parameter space.

\subsection{Solutions in MSSM parameter space}

\begin{figure}
\centering
\includegraphics[angle=90,width=8cm]{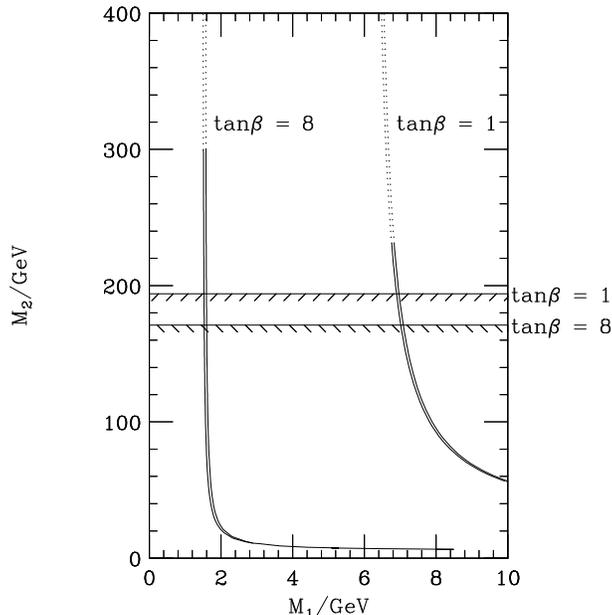}
\caption[dummy]{Solutions in $(M_1,M_2,\mu,\tan\beta)$ parameter space
 giving a $m_{\chi^0_1}=33.9$~MeV neutralino for $\mu=300$~GeV and 2
 representative values of $\tan\beta$. The width of the lines is
 0.01~MeV. Below the hashed lines the chargino mass is less than
 150~GeV. The dotted lines have $\Delta\rho_{\rm SUSY}<10^{-4}$ and
 the solid lines have $\Delta\rho_{\rm SUSY}<5 \times 10^{-4}$.}
\label{fig:mssmsols}
\end{figure}

It is important to establish whether it is indeed possible to have a
neutralino LSP with $m_{\chi^0_1}=33.9$~MeV within the MSSM.  To this
end we have scanned the MSSM parameter space
$(M_1,M_2,\mu,\tan\beta)$ with independent $M_1,\,M_2$, for a
neutralino in the mass range
\begin{equation}
 33.89 \mbox{~MeV} < m_{\chi^0_1}<33.91 \mbox{~MeV}.
\end{equation}
This leads to the neutralino iso-mass curves shown in
Fig.~\ref{fig:mssmsols}, taking $\mu=300$~GeV and 2 representative
values of $\tan\beta$. We have not been able to find any solutions
with $\mu<0$. In order to obtain such a light neutralino some
fine-tuning is required in the MSSM parameters, of about a few parts
in $10^3$ for $\tan\beta=1$ and a few parts in $10^2$ for
$\tan\beta=8$ \cite{andrea}. The fine-tuning is reduced for larger
$M_2$ and $\mu$ and small $M_1$ because a light neutralino can then be
generated by the see-saw mechanism; it is reduced for large values of
$\tan\beta$ because in the limit $\beta=\pi/2$ there is a zero mass
eigenvalue for $M_1\approx0$ \cite{andrea}.

We have checked that the Higgsino contribution always satisfies the
bound (\ref{higgs}). In order to avoid an observable light chargino we
require $m_{\chi^\pm_1}>150$~GeV, which eliminates the region below
the hashed lines in Fig.~\ref{fig:mssmsols} for the specified values
of $\tan\beta$.

The LSP is indeed dominantly bino along the solution curves in
Fig.~\ref{fig:mssmsols}. The second lightest neutralino,
${\tilde\chi}^0_2$, is dominantly wino for $M_2<300$~GeV, while for
larger values it is mainly higgsino. For $M_2\gtrsim110$~GeV, $m_
{{\tilde\chi}^0_2}\gtrsim100$~GeV, and for $M_2\gtrsim235$~GeV,
$m_{{\tilde\chi}^0_2}\gtrsim200$~GeV.

%
%

\subsection{Bounds from oblique electroweak radiative corrections}

Any new physics which couples to the SM can give contributions to the
electroweak precision observables via radiative corrections. The
effect of the new physics on vacuum polarization diagrams, the so
called oblique radiative corrections, is usually parameterised using
either the $S,\,T,$ and $U$ parameters of Ref.\cite{Peskinetc} or the
$\epsilon_1$, $\epsilon_2$ and $\epsilon_3$ parameters of
Ref.\cite{Barbieri}. The calculation of these parameters is based on
an expansion in $q^2/M_{\rm new}^2$, where $q^2$ is the momentum scale
of the gauge boson propagator, typically $M_Z^2$ or smaller and $M_{\rm
new}^2$ is the scale of the new physics, assumed to be well above
$M_Z^2$. If however there are new {\em light} particles in the
spectrum, as in the present case, these approximations are typically
insufficient and one must in general also calculate the box or vertex
corrections \cite{Barbieri2}. This full calculation is however beyond
the scope of our analysis and is not attempted here.

There is one exception however and that is the ratio of the charged to
neutral current neutrino--electron/muon scattering events --- the
$\rho$-parameter. This is defined at $q^2=0$ and the expansion is thus
trivial. In the following we limit ourselves to a calculation of the
contribution to the $\rho$ parameter from the full set of charginos
and neutralinos. The radiative correction to the $\rho$ parameter,
$\Delta\rho$, is given by the $W$ and $Z$ self energies with zero
momentum flow \footnote{See for example the third paper in
Ref.\cite{Peskinetc} for a derivation of this result.}
\begin{equation} 
 \Delta\rho = \frac{\Pi_{WW}(0)}{M_W^2} - \frac{\Pi_{ZZ}(0)}{M_Z^2} .
\end{equation}
The dominant SM contributions to $\Delta\rho$ arise from the (heavy)
top quark and the Higgs boson. Assuming the mass of the latter to be
$M_H=M_Z$, and subtracting the SM contributions we are left with the
experimental limit on new physics at the $2\sigma$ level of \cite{pdg}
\begin{equation}
 -3.7 \times 10^{-3} < \Delta \rho_{\rm new} <1.1 \times 10^{-3}.
\end{equation}
We have calculated the contribution to $\Delta\rho$, which we denote
$\Delta\rho_{\rm SUSY}$, from all charginos and neutralinos for given
parameter points $(M_1,M_2, \mu,\tan\beta)$. We find full agreement
with the results given in Ref.\cite{habarb}. We then determine
$\Delta\rho_{\rm SUSY}$ along the solution curves given in
Fig.~\ref{fig:mssmsols}. The dotted lines indicate solutions for which
$\Delta\rho_{\rm SUSY}<10^{-4}$, while the solid lines indicate
solutions for which $\Delta\rho_{\rm SUSY}<5\times10^{-4}$. Thus there
is no conflict at least with the experimental constraint on the $\rho$
parameter.

\subsection{Cosmological and astrophysical constraints}

Massive particles are expected to come into thermal equilibrium in the
early universe and their relic abundance is essentially the
equilibrium value at `freeze-out' when their self-annihilation rate
drops below the Hubble expansion rate. For the neutralino under
consideration, the self-annihilation cross-section is (s-wave)
suppressed \cite{ann} so the surviving abundance is rather high:
\begin{equation}
 m_{\tilde\chi_1^0} \left(\frac{n_{\tilde\chi_1^0}}{n_{\gamma}}\right)
  \approx 1.2 \times 10^{-2} \mbox{~GeV} 
  \left(\frac{m_{\tilde\chi_1^0}}{33.9 \mbox{~MeV}}\right)^{-2}
  \left(\frac{m_{\tilde{f}}}{150 \mbox{~GeV}}\right)^{4}\ ,
\label{cosmoabund}
\end{equation}
This energy density will be subsequently released when the neutralinos
decay and this has the potential to disrupt standard cosmology, in
particular primordial nucleosynthesis \cite{bbn}. Specifically, since
the neutralinos will be non-relativistic during nucleosynthesis, the
Hubble expansion rate will be speeded up, while the electromagnetic
energy generated through the subsequent decays will dilute the
nucleon-to-photon ratio, resulting in an increased helium-4
abundance \cite{cosmo}. The decay electrons will also Compton scatter
the thermal background photons to energies high enough to directly
alter the abundance of e.g. deuterium through photodissociation
\cite{taunu}. The observationally inferred primordial abundances thus
enable stringent bounds to be placed on the relic abundance of the
decaying particle as a function of its lifetime. For the above
abundance (\ref{cosmoabund}), the decay lifetime is required to be
less than a few thousand seconds in order that the primordial D/H
ratio is not reduced below its conservative lower limit of $10^{-5}$,
and further required to be less than a few hundred seconds in order
that the primordial $^{4}$He mass fraction not exceed its conservative
upper limit of 25\% \cite{cosmo}. Thus the cosmological lifetime bound
is essentially the same as the one derived earlier (\ref{psi4}) from
experimental considerations.

Very light neutralinos can also be produced through nucleon
bremsstrahlung and $e^+e^-$ annihilation in supernovae such as
SN~1987a. Since the squark/selectron masses are now restricted to be
above $m_W$ \cite{pdg}, the neutralinos cannot be trapped in the
supernova core by scatterings on nuclei or electrons, so will escape
freely. The energy lost through this process can be comparable to the
neutrino luminosity so may result in significant shortening of the
$\bar{\nu_e}$ burst. The neutralino luminosity can be decreased by
increasing the sfermion mass but it has been shown that consistency
with observations of SN~1987A is not possible for any sfermion mass
less than ${\cal O}(1)$~TeV \cite{snphotino}. This constraint is
evaded if the neutralino is unstable
 due to R-parity violation, as in
the present case. However there are then further constraints on the
energy released in the decays. Given the experimental upper bound
(\ref{psi4}) as well as the cosmological upper bound on the lifetime,
the decays would have occurred {\em within} the progenitor
star. Moreover the lower bound (\ref{bounds}) on the lifetime implies
that the neutralinos cannot have decayed within the supernova
core. The electromagnetic energy released in the decays would have
been thermalised leading to distortions of the lightcurve. However the
neutralinos under consideration here have a mass which is of the same
order as the core temperature \cite{raffelt} so one must reconsider
their production rate in order to quantify this potentially important
constraint.

\section{Other implications for R-parity violating phenomenology}

\subsection{The HERA high-$Q^2$ anomaly}

In 1997, the HERA collaborations reported an anomaly in their high
$Q^2$ data \cite{hera} whose most likely explanation was in terms of
R-parity violation \cite{herath}. However, this required a significant
$L_eQ_iD^c_j$ operator. While this is not excluded by our model it is
a completely distinct possibility. Together the two operators might
contribute to the decay $\mu\rightarrow e\gamma$. However this does
not lead to a significant new bound since the bound on the coupling
$\lambda'_{211}$ is already so strict \cite{huitu}.

\subsection{Neutrino masses}

The trilinear R-parity violating couplings we have introduced also
generate Majorana masses for neutrinos through one-loop self-energy
diagrams \cite{numass}:
\begin{equation}
 m_{j i} = \sum_{k, a, b} \frac{N_c}{16\pi^2}
           \frac{\lambda'_{i k b}\lambda_{j a k}}{m^2_{\tilde f}} 
           (m^2_{\rm LR})_{ab} m_{f_k},
\label{mass}
\end{equation}
where $N_c$ is a colour factor, and $(m^2_{LR})_{ab}$ is the
left-right mixing term in the sfermion sector. The question then
arises whether our model can account also for the SuperKamiokande data
suggesting oscillation of atmospheric $\nu_\mu$ into $\nu_\tau$
\cite{superk}. The results indicate that the neutrinos mix almost
maximally and that they are nearly mass-degenerate,
$\delta\,m^2\sim10^{-3}-10^{-2}$~eV$^2$. If neutrino masses are
hierarchical then the natural interpretation is that one of the
neutrinos (presumably the $\nu_\tau$) has a mass of ${\cal O}(0.1)$~eV
(although the possibility of a close mass-degeneracy for a heavier
$\nu_\mu$--$\nu_\tau$ pair is not excluded). Now the sfermion
left-right mixing is not well determined, however within a given
framework, e.g. supergravity-inspired models, approximate relations
such as $(m^2_{LR})_{aa}\approx\,m_{f_a}m_{\tilde f}$ arise. Thus
$\lambda'_{211}$, the coupling responsible in our model for the pion
decaying to the neutralino, generates a mass
\begin{equation}
 m_{\mu\mu} \approx 1.5 \times 10^{-7} \mbox{~eV} 
                    \left(\frac{\lambda'_{211}}{10^{-4}}\right)^2
                    \left(\frac{m_{\tilde{f}}}{150\mbox{~GeV}}\right)^{-1},
\label{numumass}
\end{equation}
which is too small to be of phenomenological interest. The couplings
$\lambda_{1\{2,3\}1}$ responsible for neutralino decay also generate
rather small masses:
\begin{equation}
 m_{\mu\mu,\tau\tau} \approx 10^{-6} \mbox{~eV}
                    \left(\frac{\lambda_{121,131}}{10^{-2}}\right)^2
                     \left(\frac{m_{\tilde{f}}}{150\mbox{~GeV}}\right)^{-1}.
\end{equation}
Thus the absolute scale of the masses seems too low to explain the
data on atmospheric neutrinos.  However if other R-parity violating
couplings are also present, it may well be possible to generate a
neutrino mass pattern consistent with the observations, both of
atmospheric and solar neutrinos \cite{morenu}.

\section{Future Tests}

\begin{figure}
\centering
\includegraphics[angle=90,width=0.48\textwidth]{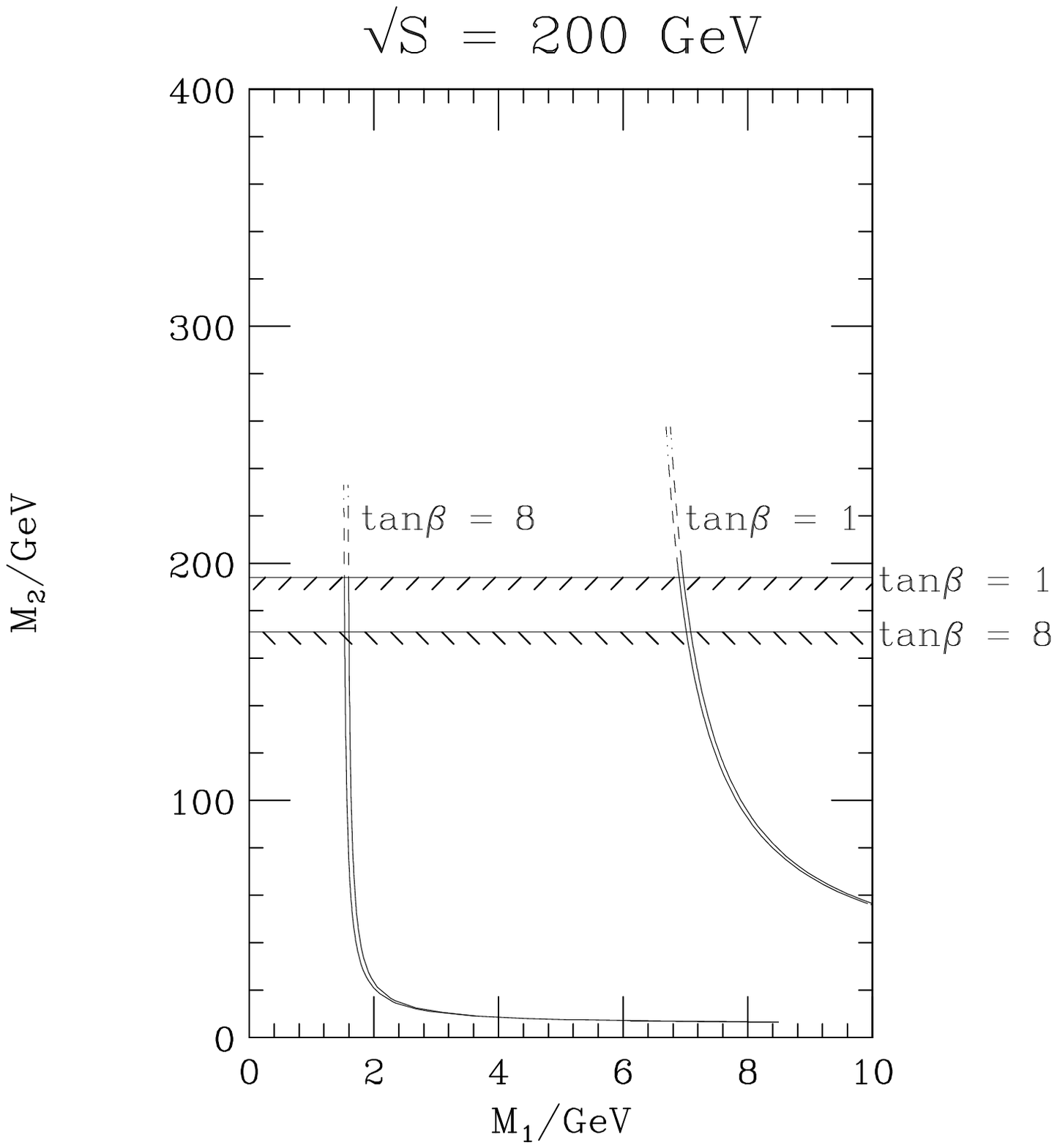}
\hfill
\includegraphics[angle=90,width=0.48\textwidth]{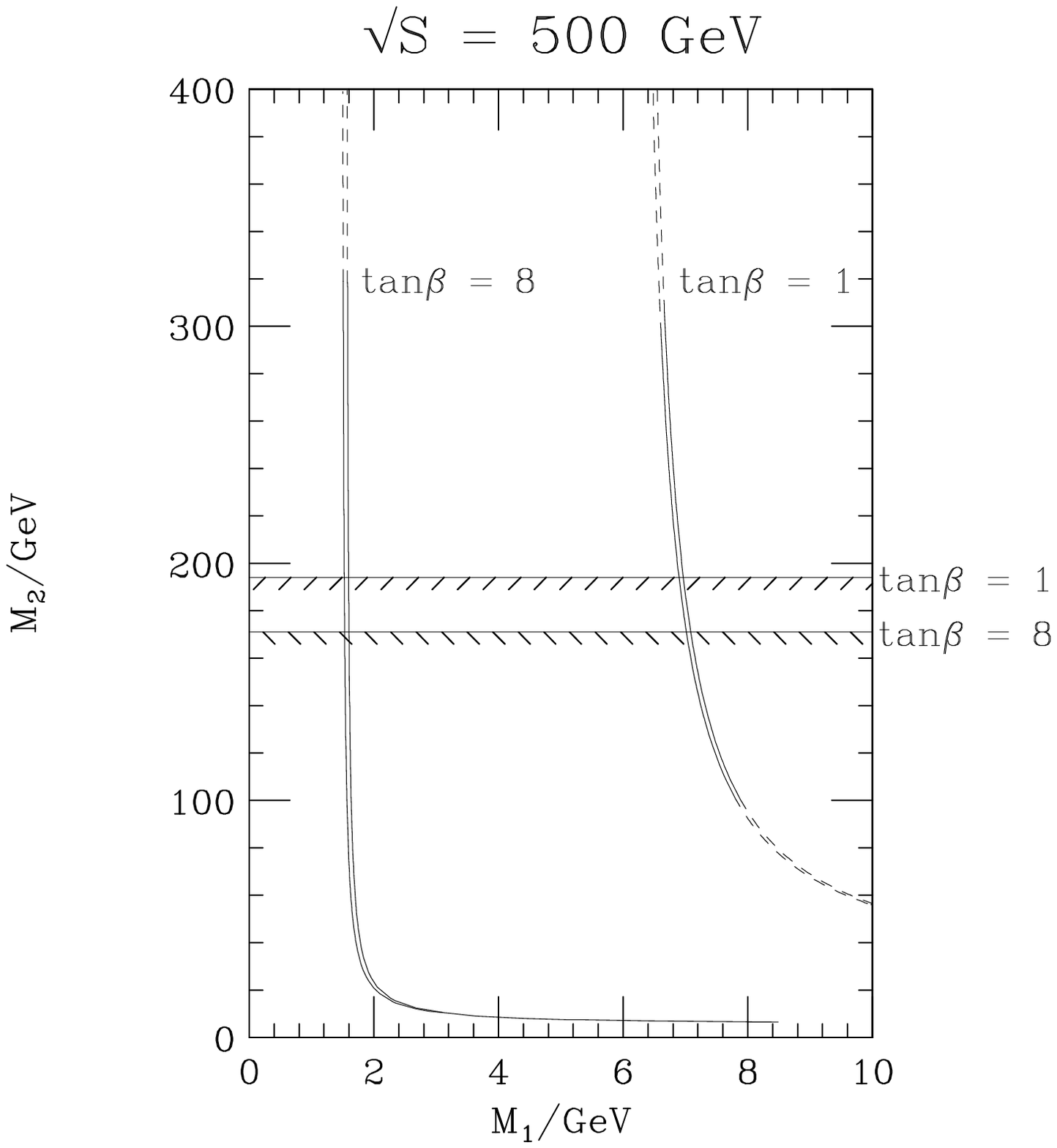}
\caption[dummy]{Cross-Sections for $e^+e^-\rightarrow
 {\tilde\chi}^0_2{\tilde\chi}^0_1$ for the solutions in
 Fig.~\ref{fig:mssmsols}. The solid lines correspond to
 0.1~pb$<\sigma<$1~pb, the dashed lines to 10~fb$<\sigma<$0.1~pb, the
 dotted lines to 1~fb$<\sigma<$10~fb, and the dot-dashed lines to
 $\sigma<$1~fb.}
\label{fig:chi2chi1}
\end{figure}

Experimentally, our model largely looks like the MSSM with
non-universal gaugino masses and with a very light LSP. Thus most
future tests of the MSSM also apply to our model, e.g. chargino pair
production. A specific test would be to identify a very light LSP for
example via neutralino pair production \cite{kneur}. At an $e^+e^-$
collider one can study the process
\begin{equation}
 e^+ + e^- \rightarrow \tilde\chi^0_2 + \tilde\chi^0_1,
\end{equation}
where ${\tilde\chi}^0_2$ subsequently decays visibly \cite{bartl}.  In
Fig.~\ref{fig:chi2chi1} we show the cross section evaluated along our
MSSM solution curves for both LEP2 ($\sqrt{s}=200\mbox{~GeV}$) and the
NLC ($\sqrt{s}=500\mbox{~GeV}$). This should be directly observable,
provided it is kinematically accessible,
i.e. $m_{\tilde\chi^0_2}<\sqrt{s}$.

%
%

Experimentally the main difference between R-parity violation with a
long-lived neutralino LSP and the MSSM is the possibility of resonant
sparticle production. The value of $\lambda'_{211}\lesssim10^{-4}$ is
too small for the observation of resonant slepton production at hadron
colliders \cite{peter}. However values of $\lambda_{1\{2,3\}1}>10^
{-3}$ should allow a test for resonant second or third generation
sneutrino production at $e^+e^-$-colliders for masses upto
$\sim\sqrt{s}$ \cite{resslep}. One can also test for the first
generation via the mechanism described in Ref.\cite{williams}.

\begin{figure}[t]
\centering
\includegraphics[angle=90,width=0.5\textwidth]{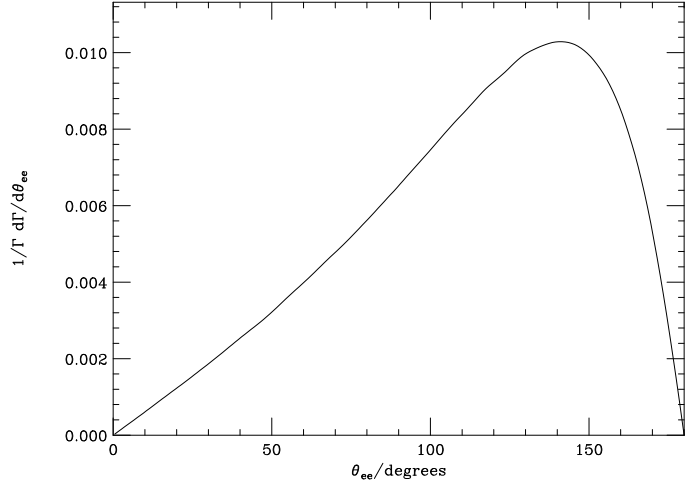}
\caption[dummy]{Decay rate of the LSP versus the angle between the final
 state electrons.}
\label{fig:angledist}
\end{figure}

A further upgrade of the KARMEN detector may allow a better resolution
of the decay of the $x$ particle, in particular the angular
distribution of the decay products. For reference we show in
Fig.~\ref{fig:angledist} the differential decay rate of the LSP in our
model as a function of the angle between the two final state
electrons.

\section{Conclusions}

The KARMEN time anomaly is particularly intriguing because contrary to
several other reported $3-4\sigma$ effects in the literature, its
significance has not diminished with improved statistics, nor has it
been explained away as a systematic effect. In fact the anomaly
persists in the KARMEN-2 data, which has a much reduced background
\cite{win}, with the same characteristics as in the KARMEN-1 data
\cite{karmen}. It would appear that there is no independent experiment
which has the sensitivity to reproduce this result. In particular
although the LSND experiment also studies pions and muons decaying at
rest, it lacks the distinctive time-structure of the beam in the
KARMEN experiment necessary to isolate the anomaly. Since KARMEN-2
acquires only of ${\cal O}(10)$ anomaly events per year of running, it
is clear that a definitive resolution of the problem will have to
await an upgraded detector with tracking capability.

Phenomenological models for the anomaly as due to the production and
decay of a new particle are very tightly constrained. The only viable
proposals at present are a singlet neutrino decaying through its large
mixing with the $\nu_\tau$ \cite{barger,govaerts}, or a neutralino
decaying through violation of R-parity \cite{choudhury} which we have
extended and investigated in detail. An important lesson from our
investigation is that contrary to popular belief a neutralino lighter
than even the pion is {\em not} excluded by present accelerator data
unless a GUT relation between gaugino masses is assumed. Whether the
KARMEN anomaly is indeed the first evidence for such a particle is a
matter for future experiments to decide.

\bigskip \noindent {\bf Note Added:} While this manuscript was under
review, the E815 (NuTeV) experiment at Fermilab reported a search for
a 33.9 MeV neutral particle produced in pion decay decaying to a
partially electomagnetic state such as $e^+e^-\nu$ or $\gamma\nu$
(J.A. Formaggio {\em et al.}, hep-ex/9912062). No evidence was found
for such a particle but the lifetimes probed ($\sim10^{-9}-10^{-3}$~s)
are much smaller than the lower limits (\ref{bounds}) on the
neutralino lifetime in our model so there are no implications. We note
however that the exclusion of such short lifetimes is relevant in the
context of the constraints from SN~1987a on the decaying particle
hypothesis \cite{barger,choudhury}. These constraints have been
investigated further in two other recent reports (I. Goldman,
R. Mohapatra and S. Nussinov, hep-ph/9912465, M. Kachelriess,
hep-ph/0001160) which conclude that our model is excluded by the
observations of SN~1987a. We reserve judgement on this issue for the
reasons mentioned earlier.

\acknowledgments{We would like to thank the following colleagues: Bill
Murray for rekindling our interest in this problem, J\"urgen
Reichenbacher and Norman Booth for details of the KARMEN~2 data, Bill
Louis for discussions on the LSND experiment, Uli Nierste and Hartmut
Wittig for clarifying the R-parity violating pion matrix element,
Andrea Romanino for his useful remark concerning fine-tuning, and
Roger Phillips for encouragement. P. Richardson acknowledges the award
of a PPARC research studentship (PPA/S/S/1997/02517).}

\appendix

\section{Calculation of the pion decay rate}

The rate of the pion decay $\pi\rightarrow\mu+{\tilde\chi}^0_1$ can be
calculated using chiral perturbation theory. To do this we need an
effective Lagrangian for the 4-fermion interaction $u$, $\bar{d}$,
${\tilde\chi}^0_1$ and $\mu^+$ with the sfermion degrees of freedom
integrated out (analogous to using the Fermi Lagrangian in the SM
calculation). This gives
\begin{equation}
 {\mathcal L} =  \sqrt{2} \lambda'_{211}
 \left(\frac{A_e}{M^2_{\tilde{\mu}}}-\frac{A_u}{2M^2_{\tilde{u}}}
 - \frac{A_d}{2M^2_{\tilde{d}}}\right) 
   \left(\bar{\mu}P_{\rm R}\tilde{\chi}_0\right)
   \left(\bar{u}P_{\rm R} d\right),
\end{equation}
where we have also Fierz-reordered the Lagrangian (and neglected some
tensor-tensor interaction terms which cannot contribute to the pion
decay rate). The matrix element of the axial-vector current between
the pion and the vacuum is
\begin{equation}
 \langle 0 |  j^{\mu 5 a}(x)|\pi^b(p)\rangle = -i p^\mu f_\pi
 \delta^{ab} {\rm e}^{-ipx},
\label{chiral}
\end{equation}
where $a$ and $b$ are isospin indices. Using Eq.~(\ref{chiral}) we
obtain
\begin{equation}
 \langle 0|\bar{u} \gamma^\mu \gamma_5 d|\pi^- \rangle =
 - i\sqrt{2}  p^\mu f_\pi {\rm e}^{-ipx}.
\label{chiral2}
\end{equation}
Contracting this with the pion 4-momentum and using the Dirac
equation for the up and down quarks yields
\begin{equation}
 \langle 0 | \bar{u} \gamma_5 d | \pi^- \rangle = 
 + \frac{i\sqrt{2} f_\pi m^2_\pi {\rm e}^{-ipx}}{(m_d + m_u)},
\end{equation}
where $m_d,\,m_u$ are thus the current quark masses.
The amplitude for the decay (\ref{piondec2}) is then,
\begin{equation}
 {\mathcal A} =  -\frac{\lambda'_{211} f_\pi m^2_\pi }{(m_d+m_u)}
 \left(\frac{A_e}{M^2_{\tilde{\mu}}}-\frac{A_u}{2M^2_{\tilde{u}}}
 - \frac{A_d}{2M^2_{\tilde{d}}}\right) 
 \left(\bar{u}_\mu(p_1)P_{\rm R} v_{\tilde{\chi}}(p_2)\right) 
  (2\pi)^4 \delta\left( p_0-p_1-p_2 \right).
\end{equation}
This gives the decay rate quoted in Eq.~(\ref{pionbr}). The additional
contribution to the decay rate $\pi^+\rightarrow e^+\nu_e$ given in
Section~\ref{sec:limits} can be calculated in the same way.

\end{document}